\documentclass[aps,10pt,reprint,groupedaddress,superscriptaddress]{revtex4-2}

\usepackage[usenames,dvipsnames]{xcolor}
\usepackage{amssymb}
\usepackage{graphicx}
\usepackage{amsmath}
\usepackage{dsfont}
\usepackage[bookmarks=true,colorlinks,citecolor=blue,urlcolor=blue]{hyperref}
\usepackage{braket}
\usepackage{blindtext}
\usepackage{physics}
\usepackage[normalem]{ulem}

\newcommand{\up}{\uparrow}
\newcommand{\dw}{\downarrow}

\newcommand{\be}{\begin{equation}}
\newcommand{\ee}{\end{equation}}
\newcommand{\fig}[1]{Fig.\thinspace{}\ref{#1}}
\newcommand{\fc}[1]{({#1})}
\newcommand{\figc}[2]{Fig.\thinspace{}\ref{#1}\thinspace{}\fc{#2}}

\begin{document}
\title{Quantum sine-Gordon Dynamics in Coupled Spin Chains}

\newcommand{\TUM}{\affiliation{Department of Physics, Technical University of Munich, 85748 Garching, Germany}}
\newcommand{\MCQST}{\affiliation{Munich Center for Quantum Science and Technology (MCQST), Schellingstr. 4, 80799 M{\"u}nchen, Germany}}

\author{Elisabeth Wybo} \TUM \MCQST
\author{Michael Knap} \TUM \MCQST
\author{Alvise Bastianello} \TUM \MCQST

\begin{abstract}
The sine-Gordon field theory emerges as the low-energy description in a wealth of quantum many-body systems. Recent efforts have been directed towards realizing quantum simulators of the model, by interfering two weakly coupled one-dimensional cold atomic gases. 
The weak interactions within the atomic clouds provide a sine-Gordon realization in the semiclassical regime. Furthermore, the complex microscopic dynamics prevents a quantitative understanding of the effective sine-Gordon validity realm.
In this work, we focus on a spin ladder realization and observe the emergent sine-Gordon dynamics deep in the quantum regime.
We use matrix-product state techniques to numerically characterize the low-energy sector of the system and compare it with the exact field theory predictions. From this comparison, we obtain quantitative boundaries for the validity of the sine-Gordon description.
We provide encompassing evidence for the emergent field theory by probing its rich spectrum and by observing the signatures of integrable dynamics in scattering events. 
\end{abstract}

\maketitle

\section{Introduction}

In recent years, highly controllable and tunable quantum simulators have been developed on different platforms, including ultracold atoms, trapped ions, and superconducting qubits. These settings enabled the experimental preparation and characterization of strongly-correlated non-equilibrium states of matter. In general, such states are extremely difficult to characterize by traditional analytical or numerical approaches, which defies a direct verification of the quantum simulation. To overcome this challenge, it is important to identify strongly correlated systems, whose dynamics can also be characterized with conventional theoretical means. In this respect, a prominent example is the sine-Gordon model \cite{coleman1975,Zamolodchikov1979,giamarchi2003quantum,mussardo2010statistical}; a one-dimensional relativistic field theory which describes the low-energy physics of a multitude of experimental systems.
The sine-Gordon model is a remarkable example of an integrable field theory \cite{Smirnov1992,Zamolodchikov1979}.
Its spectrum features topological excitations, akin to classical solitons, and their bound states, known as breathers. 
Out of equilbirium, integrability hinders thermalization and many efforts have been made to understand the exotic dynamics of this field theory \cite{Bertini2014,Kormos2016,Cubero2017,Kukuljan2018,Rylands2019,Kukuljan2020,Bertini2019}.

The sine-Gordon model captures the low-energy sector of spin chains~\cite{Affleck1999, Zvyagin2004, Umegaki2009,Tiegel2016,Bera2017,Bouillot2011}, spinful cold atom gases \cite{giamarchi2003quantum}, specific quantum circuits~\cite{Roy2019,Roy2021} and the interference of two one-dimensional quasi-condensates~\cite{Gritsev2007,Gritsev2007a}.
The latter realization of the sine-Gordon model, proposed by Gritsev \emph{et al.}~\cite{Gritsev2007}, has been realized with coupled weakly-interacting quasi-condensates on atom chips~\cite{Schweigler2017}, coupled by a potential barrier of adjustable height. The sine-Gordon Hamiltonian governs the relative phase between the two condensates, which is then probed through matter-wave interferometry
\cite{Schumm2005,Hofferberth2007}.
Importantly, the mass scale of the theory can be tuned by controlling the potential barrier.
For weak interactions within the one dimensional gases, as realized in these experiments, a semiclassical approximation of the sine-Gordon model accurately captures the equilibrium correlation functions of the relative phase \cite{Schweigler2017,Zache2020}.
\begin{figure}
\includegraphics[width=0.48\textwidth]{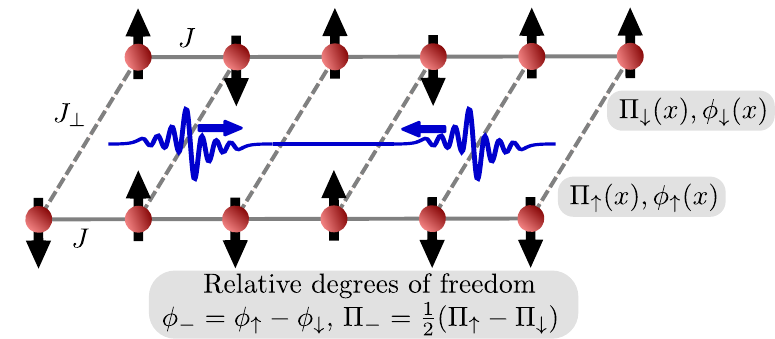}
\caption{{Sketch of the setup.} Two XXZ spin chains, each realizing a Luttinger Liquid in the low-energy regime, are weakly tunnel coupled. The sine-Gordon model is realized in their relative degrees of freedom. }\label{fig:sketch}
\end{figure}

Out of equilibrium the situation is less clear, as high-energy excitations beyond the sine-Gordon description are inevitably created. Quench experiments in tunnel coupled quasi-condensates~\cite{Pigneur2018} are not described by a semiclassical analysis of the sine-Gordon model~\cite{Horvath2019}. 
Attempts to include corrections beyond sine-Gordon arising from coupling to the symmetric-phase sector \cite{Yuri2020b} and transverse mode excitations \cite{Yuri2020a} seem not to be sufficient to quantitatively describe the experimental observations \cite{Pigneur2018}. 
From a theoretical vantage point, both the massless and large mass limits (corresponding to large and weak barrier strengths respectively) become quadratic theories, and their dynamics are readily tractable \cite{Foini2015,Foini2017,Ruggiero2021,Yuri2018}. Hence, they represent a good starting point for self-consistent Gaussian approximations \cite{Yuri2019}. Likewise, experiments focusing on these regimes \cite{Rauer2018,Langen2013,Kuhnert2013,Gring2012,Langen2015} are well captured by these approaches.
Yet, fascinating physics arises precisely in the challenging regime of intermediate mass scales and far from the semiclassical limit, in which the rich dynamics of the field theory is unveiled. 
These considerations pose two questions. First, in order to probe strong quantum effects, it will be interesting to focus on the strong-coupling regime beyond the semiclassical limit. Second, assessing the validity regime of the emergent sine-Gordon descriptions in microscopic realizations will be pertinent. 

These open questions motivate our work, in which we study the sine-Gordon dynamics emerging from the tunnel coupling between two one-dimensional XXZ spin chains, see \fig{fig:sketch}, which realizes a lattice version of the original proposal of Ref.~\cite{Gritsev2007}. Our motivation to investigate lattice systems is twofold. On the one hand, matrix-product-state techniques \cite{Verstraete2008, Schollwock2011, Hauschild2018} allow for a precise numerical characterization of the spectrum and the dynamics of the system. Therefore, well-defined boundaries to the effective sine-Gordon dynamics can be obtained.
On the other hand, interactions in these settings are strong and result in sine-Gordon realizations deep in the quantum regime, in contrast with experiments on coupled quasi-condensates.
A realization of a model closely related to the spin ladder, that consists of coupled Bose-Hubbard chains is realizable with current experimental capabilities~\cite{Bloch2008}.

Our discussion is organized as follows: in Section \ref{sec_model} we introduce the coupled spin chains and review the emergence of the effective sine-Gordon dynamics. Section \ref{sec_quantumSG} provides a short summary of the integrability aspects of the sine-Gordon model, which are then compared in Section \ref{sec_spectralF} with the numerical low-energy spectrum of the coupled chains. In Section \ref{sec_scattering} we analyze nonequilibrium scenarios and observe key signatures of integrable dynamics in scattering events of wave-packet excitations.
The appendices containing more technical aspects then follow the conclusions and outlook~\ref{sec_conclusions}.
\bigskip

\section{The sine-Gordon field theory on a spin ladder}
\label{sec_model}
The sine-Gordon model is the central character of this work and is governed by the Hamiltonian 
\be\label{eq_SG}
H_{\text{SG}}=\int \dd x \, \frac{c^2 g^2}{2} \Pi^2+\frac{1}{2g^2}(\partial_x\phi)^2- \frac{c^2m^2}{g^2} \cos(\phi)\, ,
\ee
where the bosonic fields are canonical conjugates $[\phi(x),\Pi(y)]=i\delta(x-y)$ and $\phi$ has the meaning of a phase.
The Hamiltonian is brought to the standard field theory notation \cite{mussardo2010statistical} by a simultaneous rescaling of the fields $\phi\to g\phi$ and $\Pi\to g^{-1}\Pi$, but in our context Eq. \eqref{eq_SG} is more convenient.

Above, $c$ is the light-cone velocity of correlation spreading, $m$ tunes the overall mass scale, and $g$ is the interaction.
The parameter $g$ governs the scaling dimension \cite{giamarchi2003quantum} (in an RG sense) of $\cos\phi$: for $g^2>8\pi$ the cosine term is irrelevant and the theory flows to the free boson conformal point. For $g^2<8\pi$, interactions become relevant. As a consequence, the mass parameter $m$ has an anomalous dimension scaling as $[m] = {1-g^2/(8\pi)}$. 
We postpone a more detailed discussion of the field theory to Section~\ref{sec_quantumSG}. Here, we present its realization in tunnel-coupled spin chains (see Fig. \ref{fig:sketch}) following the method of Ref. \cite{Gritsev2007}.
We thus consider two weakly tunnel-coupled XXZ spin chains 
\begin{equation} \label{eq:H}
H = H^{\mathrm{XXZ}}_{\uparrow} + H^{\mathrm{XXZ}}_{\downarrow} + H_{\perp},
\end{equation}
where
\begin{equation}\label{eq_xxz_chain}
H^{\mathrm{XXZ}}_{\alpha} = J\sum_{i=0}^{L-2} \left( S_{i\alpha}^{x} S_{i+1\alpha}^{x} +  S_{i\alpha}^{y} S_{i+1\alpha}^{y} + \Delta  S_{i\alpha}^{z} S_{i+1\alpha}^{z} \right)
\end{equation}
with $\alpha = \uparrow,\downarrow$ and $(S_{i\alpha}^{x},S_{i\alpha}^{y},S_{i\alpha}^{z})$ the spin-$1/2$ operators.
We choose the coupling between the two chains $H_\perp$ in the form of a weak tunneling
\begin{equation}\label{eq_h_perp}
H_{\perp} = \frac{J_{\perp}}{2} \sum_{i=0}^{L-1} \left( S_{i\uparrow}^{+} S_{i\downarrow}^{-} +  S_{i\uparrow}^{-} S_{i\downarrow}^{+}  \right)
\end{equation}
with $S_{i\alpha}^{\pm} = S_{i\alpha}^{x} \pm i S_{i\alpha}^{y}$. Here, the coupling $J_\perp$ is assumed to be small (specified below).
To see the emergence of sine-Gordon physics, one first neglects the tunneling and focuses on the effective low-energy description of the two chains, assuming to be close to the ground state.
To this end, one proceeds within the Luttinger Liquid approach \cite{Haldane1981,giamarchi2003quantum,Cazalilla2004} by introducing a phase field $\phi_\alpha$ and its conjugate field $\Pi_\alpha$ for each of the two chains.
It should be stressed that the Luttinger Liquid approach is of much wider applicability than the spin chain and only requires a $U(1)$ conserved charge (the $z-$magnetization in this case) and gapless excitations.
In order to have linearly-dispersing gapless excitations over the ground state of Eq. \eqref{eq_xxz_chain}, we focus on the XY-phase with $\Delta\in[-1,1]$.

Each spin chain is thus described by the Luttinger Liquid Hamiltonian
\begin{equation}
H^{\mathrm{LL}}_{\alpha}=\frac{1}{2\pi}\int \dd x v_s \left(\frac{1}{K} (\pi \Pi_{\alpha}(x))^2+K(\partial_x \phi_{\alpha}(x))^2\right)\, ,
\end{equation}
where the Luttinger parameter $K$ and sound velocity $v_s$ fully characterize the many-body interactions.
Within the bosonization language, the spin operators are represented as \cite{lukyanov1998}
\begin{equation}\label{eq_spin_bosonization}
S_j^z\simeq \Pi(x)\hspace{2pc} S_j^+ \simeq \alpha e^{i\phi(x)}\, ,
\end{equation}
where only the most relevant terms (in an RG sense) are retained.
The constant prefactor $\alpha$ in front of the phase is non-universal and depends on the microscopic properties of the model. We will treat it as a constant, that will however renormalize the bare mass of the sine-Gordon model, so it must be carefully taken into account.

As a second step, one now reintroduces the coupling between the two chains within the bosonization approach $H_\perp=|\alpha|^2 J_\perp\int \dd x  \cos(\phi_\uparrow-\phi_\downarrow)$. This strategy is only valid in the weak tunneling regime.
Finally, one rotates the fields as $\phi_{\pm}=\phi_\uparrow\pm \phi_\downarrow$ and $\Pi_{\pm}=(\Pi_\uparrow\pm \Pi_\downarrow)/2$. In this new basis, the symmetric and anti-symmetric degrees of freedom are explicitly decoupled. While the former are still described by a non-interacting Luttinger liquid, the sine-Gordon Hamiltonian naturally emerges in the anti-symmetric sector
\begin{multline}\label{eq_H_SG_micro}
H_-= \int \dd x   \frac{ v_s}{2\pi}\left(\frac{2}{K} (\pi \Pi_-)^2+ \frac{K}{2}(\partial_x \phi_-)^2\right) \\
+   J_{\perp} \int \dd x  |\alpha|^2 \cos(\phi_-).
\end{multline}
The equivalence with $H_\text{SG}$ \eqref{eq_SG} is readily established with the identification $c=v_s$, $cg^2=2\pi/K$ and $J_\perp |\alpha|^2=c^2m^2/g^2$. The sign of $J_\perp$ is unimportant, since it can be changed by a global shift $\phi_-\to \phi_-+\pi$.

Various tunnel coupled one-dimensional discrete or continuous models would equally serve our purposes for realizing the sine-Gordon model. However, considering XXZ spin chains has the great advantage that the Luttinger parameters are analytically available. Indeed, the XXZ spin chain is a well known integrable model whose exact solution in the zero magnetization sector gives \cite{Sirker2006}
\begin{equation}\label{eq:xxz_K}
K = \frac{\pi}{2} \frac{1}{\pi-\arccos\Delta}
\quad \quad 
v_s = J\frac{\pi}{2} \frac{\sqrt{1-\Delta^2}}{\arccos\Delta} \;.
\end{equation}

We tested the assumption of the decoupling between the symmetric and antisymmetric sectors by measuring correlation functions
and numerically extracting the Luttinger parameter $K$ of the symmetric sector in Appendix~\ref{app:K_Jperp}.
These estimates indicate that $K$ in the weak coupling regime $J_{\perp}/J<0.3$, is very well compatible with the single-chain exact  result, Eq.~\eqref{eq:xxz_K}, showing that hybridization effects between the symmetric and antisymmetric sectors induced by beyond-sine-Gordon corrections are small. In the remainder of this paper we will at most set $J_{\perp}/J=0.2$ to be in the sine-Gordon regime and determine $K$ and $v_s$ from Eq.~\eqref{eq:xxz_K}.

We would like to briefly comment on the asymmetric choice in the interchain coupling \eqref{eq_h_perp}: the manifest absence of a spin-spin interaction in the $z$-direction is not accidental. Indeed, such an additional term would spoil the sine-Gordon effective description, by coupling the symmetric and antisymmetric sectors in a highly non-trivial way
\cite{Schulz1986, giamarchi2003quantum}.
While unproblematic on the theoretical ground, a complete suppression of the $S^z_\uparrow S^z_\downarrow$ coupling can be challenging in experimental spin chain realizations. To this end, a more convenient experimental setup can be realized with the Bose-Hubbard model with on-site interactions and nearest-neighbor hoppings \cite{Bloch2008}. However, for our numerical studies the restricted local Hilbert space of the spin chain and the analytic solutions for its Luttinger parameters are advantageous over directly simulating the Bose-Hubbard model, which is why we focus on the former.

\section{The quantum sine-Gordon model}
\label{sec_quantumSG}
\begin{figure*}[t!]
\centering
\includegraphics[width=0.99\textwidth]{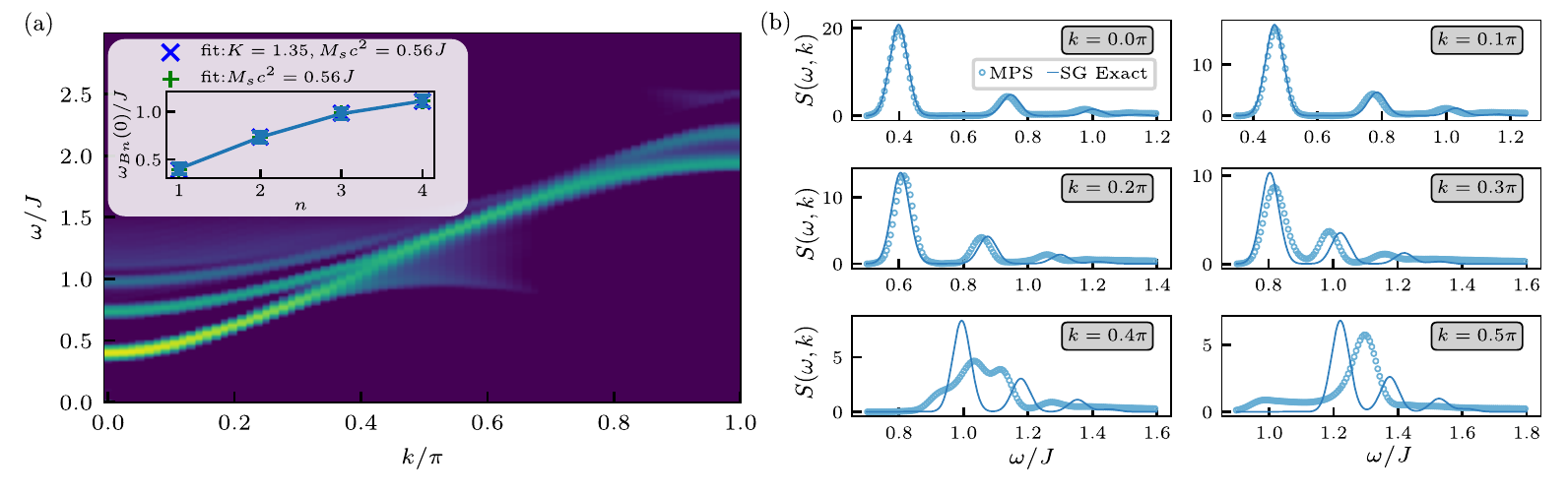}
\caption{(a) The spectral function for a system of $L=151$ rungs, with model parameters $\Delta=-0.4$ and $J_{\perp} =0.2J$. We observe $N=4$ sine-Gordon breathers as expected for $K=1.35$ of the single chain. The color scale is logarithmic in the spectral weight. The inset shows a fit of the breather energies $\omega(k=0)=c^2M_{B_n}$ according to Eq.~\eqref{eq:sg_mass}. The blue crosses are obtained by using both $K$ and $c^2M_s$ as fitting parameters, while the green pluses are obtained from $c^2M_s $ as only fitting parameter, illustrating the validity of the single-chain analytic result. (b) Momentum slices of the numerical spectral function compared to the analytical result Eq.~\eqref{eq:spectral_f_SG}, where only single breather contributions are included. At high momenta there are deviations from the analytic result.}
\label{fig:spectrum_Delta_-04}
\end{figure*}
We briefly review the main features of the sine-Gordon field theory and its exact solution, for an exhaustive discussion see Refs. \cite{Smirnov1992,mussardo2010statistical}.
We mostly refer to the notation of Eq. \eqref{eq_SG}, but occasionally relate $g$ and $K$ explicitly when convenient.
As already mentioned in the introduction, the sine-Gordon field theory belongs to the class of exactly solvable models: its integrability is established both on the classical \cite{novikov1984} and the quantum level \cite{Zamolodchikov1979}.
The key characteristic of integrable systems is the presence of infinitely many local conservation laws \cite{korepin1997}: this has deep consequences on the excitation spectrum and dynamics.
First, conservation laws ensure the existence of asymptotic multiparticle states, in spite of the strong interactions.
In the sine-Gordon case, the fundamental excitations are solitons and antisolitons connecting between the valleys of the cosine potential. These excitations have equal masses $M_s$ and relativistic energy $E_s(\theta)=c^2 M_s\cosh\theta$ and momentum $P_s(\theta)=c M_s \sinh\theta$, with $\theta$ the relativistic rapidity. The soliton mass scale has a highly non-trivial dependence on the interaction and bare mass scale $m$ \cite{Zamolodchikov1995}
\be \label{eq:soliton_mass}
c^2 M_s=\left(\frac{c^3m^2}{2g^2}\frac{\pi\Gamma(1/(1+\xi))}{\Gamma(\xi/(1+\xi))}\right)^{(1+\xi)/2}\frac{2\Gamma(\xi/2)}{\sqrt{\pi}\Gamma((1+\xi)/2)}
\ee
where $\xi=(8\pi/g^2-1)^{-1}=(4K-1)^{-1}$, and $\Gamma$ is the Euler-Gamma function.
In addition to solitons, the model also features non-topological excitations called breathers. Their masses are quantized according to
\be \label{eq:sg_mass}
M_{B_n}= 2M_s \sin\left( n\xi\frac{\pi}{2}\right) \, ,
\ee
where $n=\{1,...,N\}$ with $N=\lfloor\xi^{-1}\rfloor=\lfloor 4K-1\rfloor$.
Notice that the interaction $g$, or equivalently the Luttinger parameter $K$, tunes the quantumness of the model: for large values of $K$, the mass gaps between breathers diminish and ultimately merge in a continuum, which corresponds to the excitations of the classical theory. Hence, quantum effects are most prominent for small values of $K$ (i.e. strong interactions in the one-dimensional channels).

Interactions among sine-Gordon excitations are encoded in highly non-trivial scattering matrices.
Importantly, the conservation of infinitely many charges allows only for elastic scattering. In this respect, two excitations with different masses (e.g. two breathers) can only be transmitted and never reflected: in this case, the logarithmic derivative of the scattering matrix is readily connected with the post-scattering space displacement experienced by two colliding wave packets. In Section~\ref{sec_scattering} we will explicitly study such scattering events. The two-particle scattering matrix is the backbone of the integrability approach to the model and it is exactly known \cite{Zamolodchikov1979}: for the sake of completeness, we report the lengthy expression and the space-displacement interpretation in Appendix \ref{app_scatmat}.

The important task of connecting the asymptotic states with actual observables requires the knowledge of the matrix elements, also known as form factors. The form factor boostrap \cite{Smirnov1992} exploits the analytical properties of the scattering data to compute the sought-after matrix elements: this procedure is extremely challenging, but when possible leads to exact results.
For our purposes, we are mostly interested in the vertex operator $e^{i\phi_-}$
 \cite{Lukyanov1997b,Takacs2010,Koubek1993}. Here we focus on the matrix elements between the ground state $|0\rangle$ and one-breather excitations $|B_n(\theta)\rangle$ 
\be\label{eq_form_factors}
\langle 0|e^{i\phi_-}|B_n(\theta)\rangle=\mathcal{G}_g\frac{\sqrt{2}\cot(\xi\pi/2)\sin(n\pi\xi)\exp[I_n]e^{i\pi n/2}}{\sqrt{\cot(n\pi\xi/2)\prod_{s=1}^{n-1}\cot^2(s\pi\xi/2)}}
\ee
where
\be
I_n=\int_0^\infty \frac{\dd t}{t} \frac{\sinh^2(t n\xi)\sinh[t(\xi-1)]}{\sinh(2t)\cosh(t)\sinh(t\xi)}\, 
\ee
and the ground state expectation value $\mathcal{G}_g=\langle 0|e^{i\phi_-}|0\rangle$ is exactly known \cite{Lukyanov1997}
\be
\mathcal{G}_g=\left(\frac{c^3 m^2}{2g^2}\right)^{\xi} \frac{(1+\xi)\tan\left[\frac{\pi\xi}{2}\right]\Gamma^2[\frac{\xi}{2}]}{2\pi\Gamma^2\left[\frac{1+\xi}{2}\right]}\left(\pi \frac{\Gamma[\frac{1}{1+\xi}]}{\Gamma[\frac{\xi}{1+\xi}]}\right)^{1+\xi}\, .
\ee
In the next section, we begin to explore the low-energy phase space of the spin ladder and quantitatively compare it with the sine-Gordon predictions.

\section{Spectral analysis of the low energy sector}
\label{sec_spectralF}

We systematically probe the low-energy sector of the spin ladder numerically and aim for a quantitative comparison with the sine-Gordon results. This allows us to provide bounds on the validity of the effective low-energy description.
As the main probe of the excitation spectrum we consider the spectral function 
\begin{equation} \label{eq:spectral_f}
S(\omega,k) = \int \dd t \; W_{\sigma}(t) e^{i\omega t}  \sum_{j=-L/2}^{L/2} e^{ikj} \ev{O_j^{\dag}(t)O_0}{0}_c\, ,
\end{equation} 
where $O_j$ is a local operator centered on site $j$, $\ket{0}$ is the ground state of the model~\eqref{eq:H}. The Gaussian envelope $W_{\sigma}(t) = \exp(-\frac{1}{2}\frac{t^2}{\sigma^2})$ smoothens out the Gibbs phenomenon in the Fourier transformation caused by the finite-time cutoff present in numerical simulations.
\begin{figure*}[t]
\centering
\includegraphics[width=0.97\textwidth]{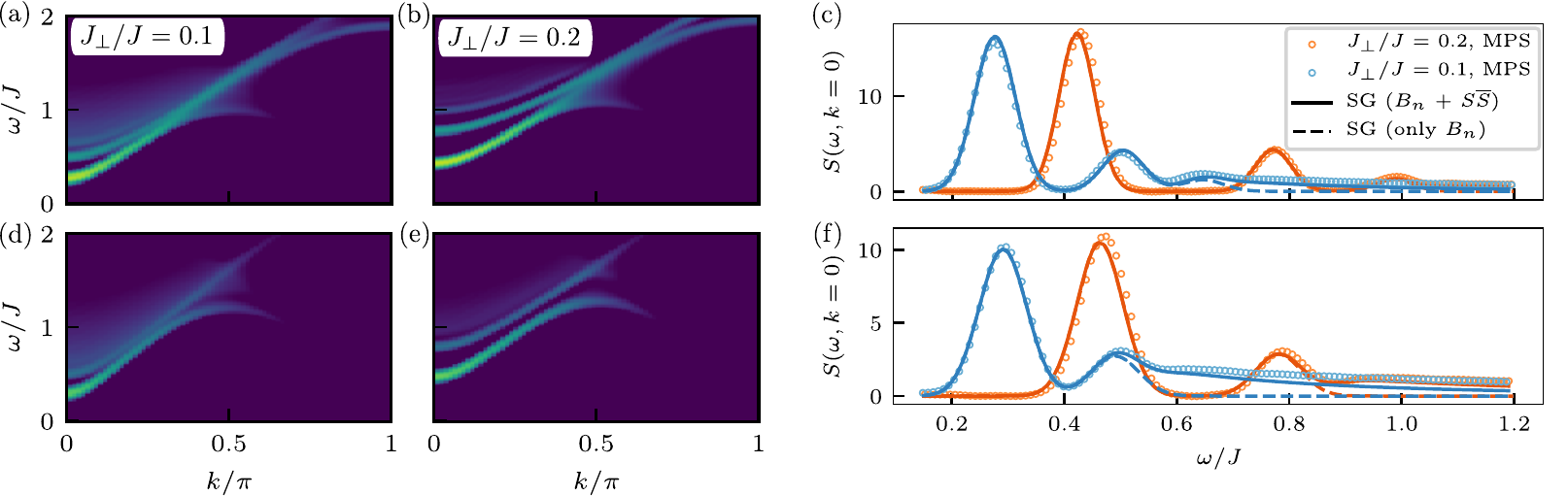}
\caption{Spectral functions for a system of $L=151$ rungs for two different values of $\Delta$ and two coupling strengths $J_{\perp}/J =0.1,0.2$. (a-c) Here $\Delta=-0.25$, corresponding to $K\simeq 1.18$ and $N=3$ breathers. (e-f) Here $\Delta=0.1$, corresponding to $K\simeq 0.94$ and $N=2$ breathers. The color scale used in the spectral-density plots is logarithmic in order to make the higher lying breathers also visible compared to the lower ones. In panels (c) and (f) we show the profile of the spectral function at $k=0$ for both coupling strengths and compare to the exact result: the solid lines take into account the breathers and the soliton/anti-soliton continuum, while the dashed lines only takes into account single breather contributions.
}\label{fig:spectra}
\end{figure*} 
To subtract the zero-frequency signal of the ground state, we consider the connected correlation function $\ev{O_j^{\dag}(t)O_0}{0}_c=\ev{O_j^{\dag}(t)O_0}{0}-\langle 0|O_j^{\dag}(t)|0\rangle\langle 0|O_0 |0\rangle$.

In order to probe the particle content of the theory, we evaluate the spectral function for the single-rung operator $O_j = S_{j\uparrow}^{+} S_{j\downarrow}^{-}\simeq |\alpha|^2 e^{i\phi_-}$, see Eq.~\eqref{eq_spin_bosonization}. The non-universal prefactor $\alpha$ is the same governing the bare mass scale \eqref{eq_H_SG_micro} and does not depend on the chain coupling $J_\perp$.
The spectral function \eqref{eq:spectral_f} can be computed by inserting a Lehmann decomposition of the identity in between the two observables and by expanding in the basis of the asymptotic states of the field theory.
By retaining only the one-particle contributions, one obtains
\begin{multline}\label{eq:spectral_f_SG}
S(\omega,k)=\sum_n \int \dd\theta \delta(k-c M_{B_n}\sinh(\theta))  \sqrt{2\pi}|\alpha|^2\sigma  \\ \times e^{-\frac{\sigma^2}{2}(\omega-c^2 M_{B_n} \cosh\theta)^2} |\langle 0|e^{i\phi_-}|B_n(\theta)\rangle|^2 \, .
\end{multline}

As a first comparison, we track the relativistic dispersion law of the breathers and the mass ratios.
To this end, we access the microscopic spectral function with matrix-product-state (MPS) techniques to evaluate the real-time correlation function~\cite{White2004,White2008}: first, we obtain the ground state $\ket{0}$ by the density-matrix-renormalization-group (DMRG) algorithm~\cite{White192,Schollwock2011}, then we perform a real-time evolution of the state $O_0\ket{0}$ with the time-dependent-block-decimation (TEBD) algorithm ~\cite{Vidal2004,White2004,Daley2004}. For further details about the numerical evaluation of the spectral function, we refer to the Appendix~\ref{app:det_spectral}. 

In \figc{fig:spectrum_Delta_-04}{a} we show the spectral function of a ladder with $\Delta=-0.4$, corresponding to $K=1.35$, and $J_{\perp} = 0.2J$. For this choice of parameters, we expect $N=4$ breathers to appear in the spectral function as can be indeed observed. In Fig.~\ref{fig:spectra} further spectral functions for a different number of breathers $N=2,3$ are shown and the coupling strength $J_{\perp}/J = 0.1,0.2$ is varied. Notice that, at least for small momenta, the signal nicely follows the expected behavior of the relativistic dispersion law $E_{B_n}(k)= \sqrt{c^4 M_{B_n}^2+c^2k^2}$. The feature observed around $k\simeq \pi/2$ is a clear deviation from the sine-Gordon description and is a remnant of the single-chain spinon dispersion.
In the single chain the lower edge of the spinon continuum at $k=\pi/2$ is perturbatively given by $(1+2\Delta/\pi)J$
~\cite{Kolezhuk1999,QuantumMagnetism2004}. The weak tunnel coupling is only expected to  strongly modify the low-energy regime $\omega \lesssim J$. In the higher energy regime our correlator perturbatively contains terms of the form $\ev{S_{j\uparrow}^{-}(t)S_{0\uparrow}^{+}}{0_{\uparrow}}$ with $\ket{0_{\uparrow}}$ the ground state restricted to the single chain, and unveils the lower edge of the spinon continuum at high momenta.

To test the validity of the mass prediction Eq.~\eqref{eq:sg_mass}, we then extract the breathers energies from the peaks of the spectral function at $k=0$.
In this case, we use the soliton energy and $K$ as fitting parameters; see \figc{fig:spectrum_Delta_-04}{a} inset.
As best fit parameters we obtain $K=1.35\pm 0.03$ and $c^2M_s/J=0.56\pm 0.01$.
The best-fit $K$ is thus compatible with the single-chain exact result of $K=1.35$ given by Eq.~\eqref{eq:xxz_K}.
The excellent agreement of numerical data with the mass law is already a clear signature of the sine-Gordon effective theory, but we now want to quantitatively compare the whole spectral functions Eqs. \eqref{eq:spectral_f} and \eqref{eq:spectral_f_SG}, thus confirming that breathers contribute through the form factor expression \eqref{eq_form_factors}.

\begin{figure}[t!]
\centering
\includegraphics[width=0.495\textwidth]{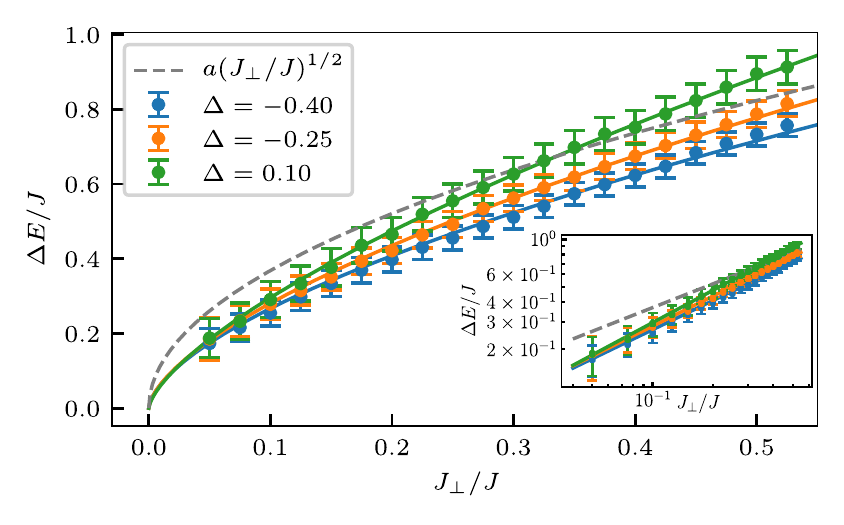}
\caption{The energy gap in the anti-symmetric sector as obtained from the dynamic correlation functions of the ground state as a function of the interchain coupling. The solid lines show a fit of the data according to $a (J_\perp/J)^{\frac{1+\xi}{2}}$ with $\xi=(4K-1)^{-1}$ fixed. The errorbars are set by the resolution in frequency space. The inset shows the same data in a logarithmic scale.}  \label{fig:K_gaps}
\end{figure}

The sine-Gordon model determines the spectral function up to an overall constant $\alpha$ coming from the single-chain bosonization \eqref{eq_spin_bosonization} which, in turn, sets the sine-Gordon bare mass scale and eventually the soliton mass through Eq. \eqref{eq:soliton_mass}. Reversing the logic, we can look at the physical soliton mass as the free parameter to be determined, which unambiguously fixes the microscopic non-universal proportionality factor $\alpha$. We first check the non-trivial dependence of the mass gap (mass of the lightest breather) with the coupling across the rungs $J_\perp$. 
Indeed, by feeding the microscopic value of the bare mass in Eq.~\eqref{eq:soliton_mass}, we obtain $\Delta E=c^2 M_{B_1}\propto J_\perp^{(1+\xi)/2}$, where the proportionality constants are $J_\perp$ independent. In Fig.~\ref{fig:K_gaps}, we show the numerically computed energy gaps as a function of the rung coupling for several interaction strengths $\Delta$, finding an excellent agreement with the non-trivial power law behavior in the whole range of explored parameters. 

Equipped with this benchmark of Eq.~\eqref{eq:soliton_mass}, we can now pursue the program outlined above and extract the parameter $\alpha$ from the measurement of the energy gap.
In \figc{fig:spectrum_Delta_-04}{b} we quantitatively compare the energy dependence of the spectral function \eqref{eq:spectral_f} with the sine-Gordon prediction \eqref{eq:spectral_f_SG} for different values of the momenta $k$: at small momenta, the agreement is excellent. We stress once again that $\alpha$ is the only fitting parameter. The Gaussian broadening of the peaks governed by $\sigma$ in Eqs.~\eqref{eq:spectral_f} and \eqref{eq:spectral_f_SG} has been set equal in both cases. As one moves to higher momenta, small deviations from the field theory prediction become evident, but the peaks of the excitations are still clearly seen (see e.g. $k=0.3\pi$). In this regime, we interpret the deviation as lattice corrections to the sine-Gordon relativistic dispersion relation, which is slightly bend towards smaller energies.
As the momentum is further increased, the spinon contribution of a single chain starts to become dominant and the energy peaks of the sine-Gordon breathers disappear.

We further analyze the spectral function for different parameters, changing the number of breathers in the sine-Gordon spectrum, in Fig.~\ref{fig:spectra}. To quantitatively probe the field theory, we show the comparison between the numerical and analytic spectral function at zero momentum $k=0$.
Reducing the value of $J_\perp$, the energy scale of the excitations is lowered and the spectral function develops a continuous tail associated with multiparticle contributions, to be contrasted with the isolated peaks of single-breather terms.
To capture this continuum, we add the soliton-antisoliton contribution to the spectral function \eqref{eq:spectral_f_SG}. The solitonic form factors of the vertex operators have been computed in Ref. \cite{Lukyanov1997b}. In Fig.~\ref{fig:spectra} we compare the spectral function at $k=0$ obtained by considering only the contribution of breathers (dashed line) with the extension to the soliton-antisoliton corrections (solid line) and the numerics (circles). The comparison shows that solitons capture the tails of the spectral function well. Depending on the value of $K$, the energy of the soliton-antisoliton pair can be comparable to the excitations of two breathers $B_1$: we checked this correction to the spectral function, but the $B_1-B_1$ contribution turned out to be negligible. We thus attribute the small discrepancy with the numerical data to corrections beyond sine-Gordon.

From the spectral functions (shown in Figs.~\ref{fig:spectrum_Delta_-04} and~\ref{fig:spectra}) it becomes apparent that the spectral weight of the higher lying breathers, with breather index $n>2$ is strongly suppressed with respect to the lower lying ones. In Appendix~\ref{app:suppr_B1} we provide a form factor argument to increase the relative weight of the higher lying breather, that is based on computing the spectral function over a correlation function of a linear combination of single- and two-rung operators.

\begin{figure}
\centering
\includegraphics[width=0.48\textwidth]{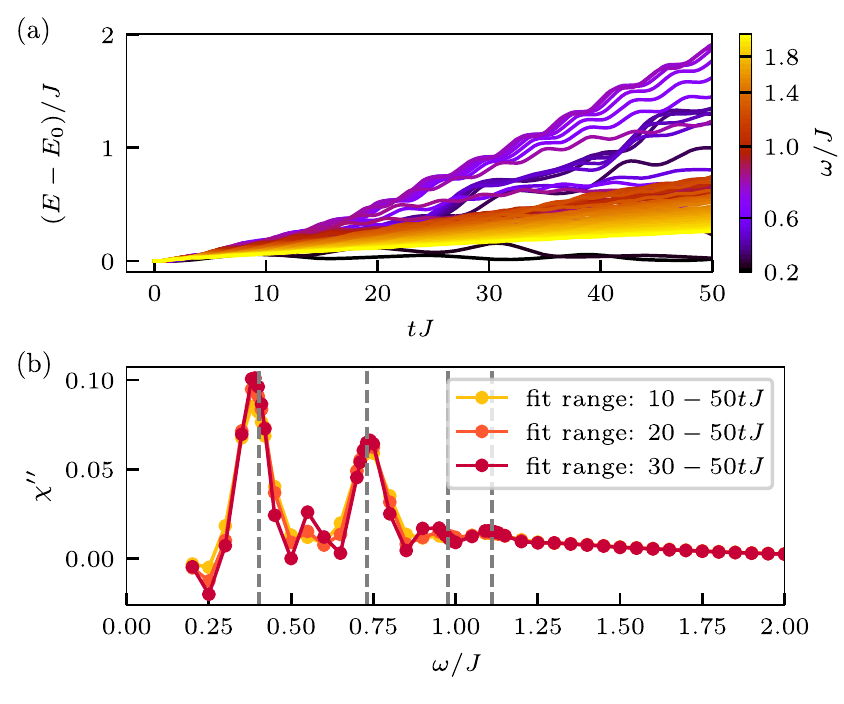}
\caption{(a) Energy absorption for a small periodic drive. Here $\epsilon=J_{\perp}/4$. The system size is $L=51$ and the model parameters are $\Delta=-0.4$ and $J_{\perp}/J=0.2$. (b) Spectral functions extracted from the absorbed power from Eq.~\eqref{eq:energy_lin_resp}. The fit of the slope is performed in several time domains (legend). The dashed grey lines show the breather energies extracted from the spectral function at $k=0$ (c.f. \fig{fig:spectrum_Delta_-04}). }\label{fig:e_resp}
\end{figure}

While the spectral function \eqref{eq:spectral_f} gives direct access to the low energy spectrum of the coupled chains, measuring unequal-time correlation functions is challenging in realistic experimental scenarios.
To this end, the fluctuation-dissipation theorem can be used to relate the spectral function to the energy absorption due to periodic drives.
We consider a periodic modulation of the Hamiltonian $H(t)=H+V(t)$, where $H$ is given by Eq.~\eqref{eq:H}, and
\begin{equation}\label{eq:p_drive}
V(t) = \epsilon\sin(\omega t) O 
\end{equation}
where we chose the perturbation as a linear combination of two terms $O=a O_1+b O_2$, with $O_1 =\sum_{i=0}^{L-1} \left(S_{i\uparrow}^{z}- S_{i\downarrow}^{z} \right)$ and $O_2 = \sum_{i=0}^{L-1} \left(  S_{i\uparrow}^{+} S_{i\downarrow}^{-} +  \mathrm{h.c.}  \right)$.

The choice of the drive is motivated on physical grounds. $O_1$ is the $z-$magnetization difference between the two chains. Hence, driving $O_1$ can be achieved by applying a small magnetic field of opposite sign to the two chains. On the other hand, modulations coupled to $O_2$ are obtained by manipulating the interchain coupling \eqref{eq_h_perp}.
Here, we focus on the linear response on the top of the ground state. Hence we limit ourselves to the case $\epsilon<J_\perp$. 
Within bosonization \eqref{eq_spin_bosonization}, $O_1\sim \int\dd x\, \Pi_-(x)$ and $O_2\sim\int\dd x\, \cos(\phi_-)$. Therefore, the operators $O_1$ and $O_2$ respectively excite odd and even breathers from the ground state.
In linear response theory, the fluctuation-dissipation theorem links the averaged absorbed energy to the spectral function at zero momentum.
Defining $E(t)=\ev{H(t)}$, one has
\be\label{eq:energy_lin_resp}
\overline{\dv{E}{t}} = \frac{1}{T} \int_0^{T} \dd t  \dv{E(t)}{t}=  \frac{\epsilon^2\omega }{4}S(\omega,0)\, ,
\ee
where $T=2\pi/\omega$ is the driving period. Hence the spectral function is directly probed by the slope of the on-average linear growth of the energy. 
Notice that the different parity of the operators $O_{1,2}$ leads to a splitting of the spectral function $S(\omega,k)=a^2 S_{O_1}(\omega,k)+b^2 S_{O_2}(\omega,k)$, where $S_{O_\mu}$ is the spectral function of the operator $O_\mu$. Hence, the choice of the coefficients $a,b$ does not have any physical meaning within the linear regime and we  consider $a=b=1$ to probe the even and odd sectors within a single measurement.

In \figc{fig:e_resp}{a} we show the energy growth for various driving frequencies, and in \figc{fig:e_resp}{b} we show the corresponding response function extracted from linear fits in several time windows. We compare the results to the breather energies extracted from the two-point correlation function, illustrating a very good agreement. The spectral weight of $B_3$ and $B_4$ is probably too low to render these excitations observable in experiments, however the ratio between $B_2$ and $B_1$ is already sufficient to obtain the mass scaling.

\section{Real-time scattering of sine-Gordon excitations}
\label{sec_scattering}
\begin{figure*}[t!]
\centering
\includegraphics[width=0.95\textwidth]{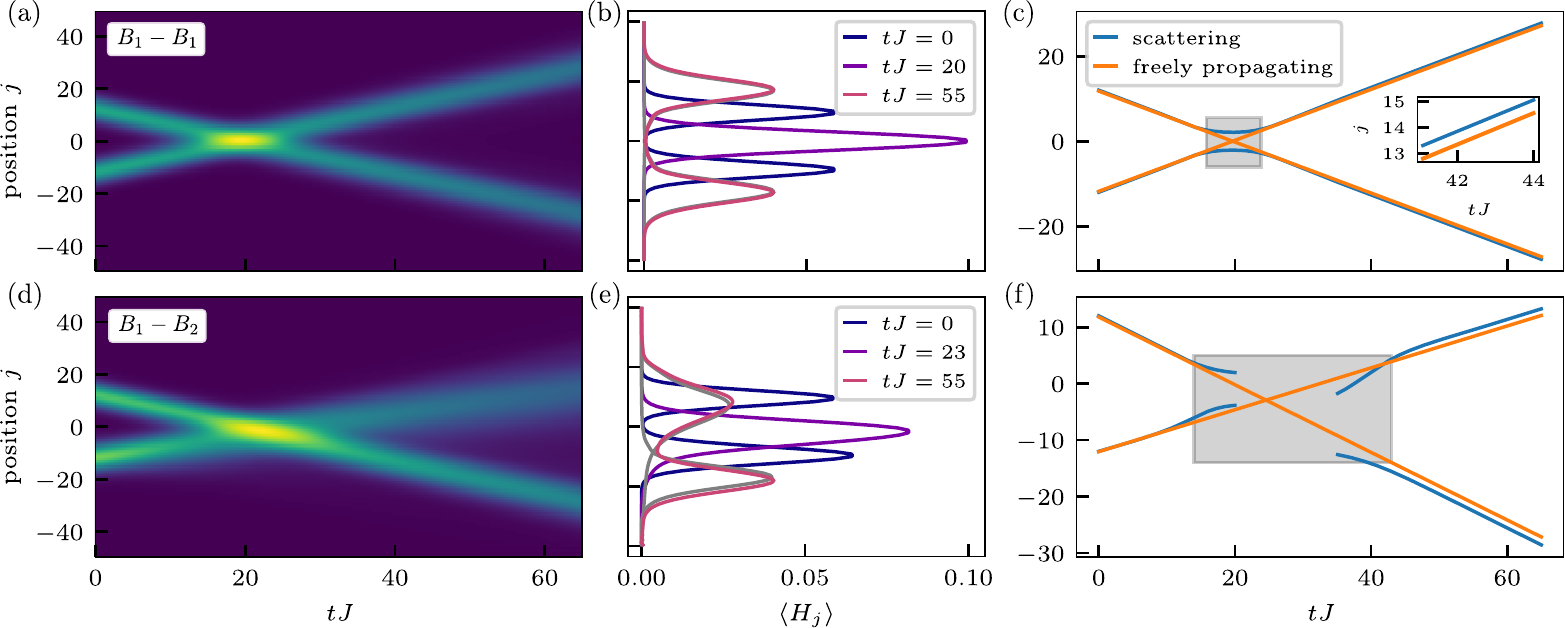}
\caption{Wave packet scattering in the low-energy regime of the coupled chains with parameters $J_{\perp}/J=0.1$, $\Delta=-0.4$ and $L=101$.  (a,d) The local energy as function of time for a scattering event between two $B_1$ quasi particles with momentum $k=\pm0.2\pi$, panel (a), and between $B_1$ and $B_2$ quasi particles with respective momenta $k_1 = -0.2\pi$ and $k_2=0.15\pi$, panel (d).
(b,e) The corresponding energy profiles at selected times, where the grey lines show the profile of the freely propagating wave packets at the latest time. (c,f) The associated average positions of the quasi particles. The numerically obtained values for the scattering displacements are $\delta x^{(11)}_{B_1} \simeq 0.5$, see also the inset in panel (c), and $\delta x^{(12)}_{B_1} \simeq 1.4$, $\delta x^{(12)}_{B_2} \simeq 1.2$.
}\label{fig:wp}
\end{figure*}
The excellent comparison between the ladder spectral function and the sine-Gordon field theory allowed us to identify the regime of validity of the latter. As a next step, we probe genuine nonequilibrium dynamics of the system.
A clear signature of integrability can be found in scattering events: the existence of infinitely many conservation laws ensures complete elastic scatterings even if non-elastic scatterings are, in principle, energetically allowed.
We thus create localized wave packets of the lighter breathers and observe their scattering dynamics. In this case, the sine-Gordon scattering matrix ensures complete transimission of breather-breather scatterings (to be contrasted with soliton-antisoliton scattering which, due to additional symmetries, can also be reflective).
Complete transmissive scattering arises also in non-interacting theories, but in the sine-Gordon field theory interactions are strong and manifest as a spatial displacement after scattering. The displacement is completely determined by the exactly known scattering matrix \cite{Zamolodchikov1979}, as we briefly report in Appendix \ref{app_scatmat} (see also Refs.~\cite{Vlijm2015,VanDamme2021} for numerical computation of scattering displacement in integrable spin chains).

The sine-Gordon excitations are collective low-energy modes that are not trivially connected to the spin configurations.
To engineer a local breather excitation, we create a Gaussian wave packet with momentum $k$, centered around site $d$ with width $\bar{\sigma}$ by applying the operator
\begin{equation}\label{eq:wp_odd}
O^{(1)}_{d,k} = \frac{1}{\sqrt{2\pi}\bar{\sigma}}\sum_j e^{-\frac{(j-d)^2}{2 \bar{\sigma}^2} + ikj} \left( S_{j\uparrow}^{z} - S_{j\downarrow}^{z}\right)\, 
\end{equation}
on the ground state $O^{(1)}_{d,k}|0\rangle$.
Within bosonization $S_{j\uparrow}^{z} - S_{j\downarrow}^{z}\simeq \Pi_-$, hence the low-energy excitations are determined by the sine-Gordon form factors $\langle B_n(\theta)|\Pi_-|0\rangle$ which are exactly known \cite{Smirnov1992}, but their analytical expression is not needed for our purposes.
The $\Pi_-$ operator is odd under the $\mathbb{Z}_2$ symmetry of the sine-Gordon model $\phi_-\to-\phi_-$. Moreover, the matrix element is non vanishing only for breathers odd under parity, which correspond to odd integers $n$, and gets smaller as $n$ is increased.
Therefore, the operator \eqref{eq:wp_odd} will with highest probability couple to the lightest breather. But this is not the only contribution: heavier breathers are expected to be excited as well and, most importantly, high-energy excitations beyond the sine-Gordon description are inevitably produced.
To eliminate the unwanted contributions and target low-energy excitations within an energy interval $\Delta E$, we perform a short imaginary-time evolution $e^{-\tau H}$, $\tau \sim 1/\Delta E$. After this, we indeed obtain clean wave packets that are moving with a speed compatible with the group velocity $v_1(k)=\partial_k E_{B_1}(k)$.
For further details about the creation of the wave packets, we refer to Appendix~\ref{app:det_wp}.

With the operator $O^{(1)}_{d,k}$ Eq.~\eqref{eq:wp_odd}, we can target the first breather $B_1$, but we also want to scatter wave packets of different species to highlight the sought-after transmission.
To this end, we also create wave packets by using another operator $O^{(2)}_{d,k}$
\begin{equation}\label{eq:wp_even}
O^{(2)}_{d,k} = \frac{1}{\sqrt{2\pi}\bar{\sigma}} \sum_j e^{-\frac{(j-d)^2}{2\bar{\sigma}^2} + ikj} \left( S_{j\uparrow}^{+} S_{j\downarrow}^{-} + \mathrm{h.c. } - \ev{o_j}   \right).
\end{equation}
 In this case, we need to explicitly impose orthogonality with respect to the ground state $\ev{O^{(2)}_{d,k}}{0}=0$ by subtracting the expectation value $ \ev{o_j}= \ev{S_{j\uparrow}^{+} S_{j\downarrow}^{-} + \mathrm{h.c.} }{0}$.
In bosonization, $[S_{j\uparrow}^{+} S_{j\downarrow}^{-} + \mathrm{h.c. }]\sim \cos\phi_{-}$, which is an even operator. Hence, in the low-energy sector $O^{(2)}_{d,k}|0\rangle$ couples only to wave packets of even breathers. Therefore, by cooling down the state $O^{(2)}_{d,k}|0\rangle$ with imaginary-time evolution, we can selectively excite wave packets of the second breather $B_2$.

\subsection{$B_1-B_1$ scattering}
We start by considering a symmetric $B_1-B_1$ scattering event, where we initialize the state as $O^{(1)}_{-d,k_2}O^{(1)}_{d,k_1}\ket{0}$ with $d=12$ and $k_2=-k_1=0.2\pi$. We show the local energy profile as a function of time in Fig.~\ref{fig:wp}(a), from this we indeed observe a very coherent motion of the wave packets. In Fig.~\ref{fig:wp}(c) we plot the average position of the scattering wave packets in the upper-half ($x_{\textrm{u-h}}$) and in the lower-half ($x_{\textrm{l-h}}$) of the system as obtained by
\begin{align}\label{eq:wp_pos}
\begin{split}
x_{\textrm{u-h}} &= \frac{1}{\mathcal{N}} \sum_{j\in \textrm{u-h}} j\ev{H_j}^2 \;
\textrm{with} \quad \mathcal{N} = \sum_{j\in \textrm{u-h}} \ev{H_j}^2, \\
x_{\textrm{l-h}} &= \frac{1}{\mathcal{M}} \sum_{j\in \textrm{l-h}} j\ev{H_j}^2 \;
\textrm{with} \quad \mathcal{M} = \sum_{j\in \textrm{l-h}} \ev{H_j}^2.
\end{split}
\end{align}
Note that these measures are only valid in the asymptotic regions before and after the scattering event, where the separation between the wave packets is much larger than their typical interaction length. Comparing the average position between the interacting and freely propagating case, we find a displacement of $\delta x^{(11)}_{B_1} \simeq 0.5$ (see inset in Fig.~\ref{fig:wp}(c)). This value can be compared to the analytical result obtained from the $S$-matrix of the sine-Gordon theory. To obtain a concrete result, we have used the following: (i) the Luttinger parameter $K$ and the sound velocity $v_s=c$ as obtained from the single chain by Eq.~\eqref{eq:xxz_K}, for $\Delta=-0.4$, $K=1.35$ and $c=0.73J$, (ii) the soliton mass $M_s=\Delta E/(2c^2\sin(\pi\xi/2))$ where we determined the energy gap $\Delta E/J=0.26$ from the spectral function, (iii) the average speed of the wave packets $v=\pm 0.60J$ which we have obtained from a fit of the freely propagating wave packets and not from the original input momentum chosen in Eq.~\eqref{eq:wp_odd} as the imaginary time evolution could potentially slightly lower the initial momentum of the wave packet (see Appendix~\ref{app:det_wp}). 
Finally, when the two wave packets collide, each excitation will scatter with all the breathers contained in the other wave packet and pile up the cumulative space displacement. Hence, we need to estimate the number of excited breathers. To this end, we consider the total energy carried by each wave packet and divide by the breather dispersion law computed at the rapidity extracted from the velocity measurement. From this analysis, we found that we excite approximately one breather in each wave packet.
Using this last piece of information, we then obtain an analytic value of $\delta x^{(11)}_{B_1} = 0.44$ in good agreement with our numerical estimate.
Notice that in the case of equal particles, one cannot distinguish between a transmission or reflection event. Hence, we now consider wavepackets of different breathers.

\subsection{$B_1-B_2$ scattering}

Next, we consider a mixed $B_1-B_2$ scattering event by initializing the state as $O^{(2)}_{-d,k_2}O^{(1)}_{d,k_1}\ket{0}$ with $k_1=-0.2\pi$, $k_2=0.15\pi$ and $d=12$. The wave packet associated with $B_2$ (the heavier quasi-particle) is expected to have a smaller velocity than the $B_1$ wave packet, and therefore also a larger broadening in time. The energy as a function of time for this process is shown in Fig.~\ref{fig:wp}(d).
In Fig.~\ref{fig:wp}(f), we plot the average position of the wave packets. As the wave packet associated with $B_2$ is moving slower we extend (decrease) the upper-half (lower-half) of the system by 10 sites in the measurement of Eq.~\eqref{eq:wp_pos} ($\textrm{u-h/l-h}= 60/40$), but only after the scattering. Before the scattering we still  divide the system in halves ($\textrm{u-h/l-h}= 50/50$). To check that is indeed a valid choice, we have also fitted a bi-modal Gaussian distribution to our data before and after scattering (no such fit is possible during the scattering event). The mean values of this distribution are in good agreement with the positions obtained from Eq.~\eqref{eq:wp_pos}. From our data we extract the displacements $\delta x^{(12)}_{B_1} \simeq 1.4$ and  $\delta x^{(12)}_{B_2} \simeq 1.2$. The analytical values obtained when plugging in the velocity of the quasi particles  $v_{B_1} = -0.60J$ and $v_{B_2} = 0.37J$, obtained from a fit to the freely propagating wave packets, are $\delta x^{(12)}_{B_1} = 1.49$ and  $\delta x^{(12)}_{B_2} = 1.18$ again in very good agreement with our numerical estimates.  In this case, we have also checked explicitly that the asymptotic velocities of the wave packets after scattering are not compatible with a reflection event, i.e. with the non-transmissive solution of the energy-momentum conservation $E_{1}(k_1)+E_2(k_2)=E_1(k_1')+E_2(k_2')$ and $k_1+k_2=k_1'+k_2'$. Indeed, plugging in the above-mentioned velocities would give $v'_{1} = 0.55J$ and $v'_{2} = -0.45J$ as velocities after reflection. This is in contrast to the $B_1-B_1$ event, for which reflection and transmission are degenerate.

\section{Conclusions \& Outlook \label{sec_conclusions}}

The sine-Gordon field theory is a ubiquitous low-energy approximation of one-dimensional quantum systems. Recent experiments focused on implementing it on highly-controllable quantum simulators based on coupled one-dimensional quasi-condensates~\cite{Gritsev2007, Schweigler2017}.
In our work, we considered two tunnel-coupled spin chains, realizing a lattice version of the sine-Gordon model proposed in Ref.~\cite{Gritsev2007}.
The lattice realization has advantages compared with the quasi-condensates: first, the one-dimensional chains can be strongly interacting, leading to a sine-Gordon model deep in the quantum regime in contrast to the weakly interacting quasi-condensates. Second, the dynamics of coupled chains can be accurately determined numerically, which allows for a direct observation of the emergent sine-Gordon description and for a quantitative analysis of its validity regime.

By a numerical characterization of the spectral function, we have studied the sine-Gordon spectrum in the low-energy sector of the chain. Comparing the numerical data with exact predictions, we obtain precise regimes for the validity of the field theory.
The observation of the rich sine-Gordon spectrum already supports the field theoretic approximation. Yet, a striking feature of the model is its integrability, resulting in unconventional nonequilibrium properties. 
As a probe of integrable dynamics, we studied scattering events of low-energy wave packets. Not only we observe fully transmissive scattering as expected from integrability, but we also indirectly observed the sine-Gordon scattering matrix by measuring the wave packet displacement after scattering.

Our work demonstrates that coupled ladder geometries are useful to realize a simulator of the sine-Gordon model in the quantum regime. Similar realizations as the one considered here based on coupled Bose-Hubbard models can be realized with current experimental techniques \cite{Bloch2008}, leading to various interesting questions for future investigations.
First, a characterization of experimentally-feasible protocols to selectively create and manipulate the sine-Gordon excitations will be of interest. To this end, numerical benchmarks will play an important role to assessing the validity of the quantum simulation and to keep unwanted effects under control.
Second, the important question of characterizing the sine-Gordon dynamics beyond the few-excitation regime presents a formidable challenge. To this end, the theory of Generalized Hydrodynamics developed for integrable models \cite{alvaredo2016,bertini2016,bastianello2022} promises access to large scale dynamics. 
Numerical benchmarks will be helpful in testing the hydrodynamics predictions on small scales and in characterizing the short-time transients that are beyond the hydrodynamic approximation.

\section{Acknowledgement}
We are grateful to F. Essler for helpful remarks on the bosonization of the coupled spin chains, and thank I. Lovas, L. Vanderstraeten and W. Kadow for useful discussions.
Tensor-network calculations were performed using the TeNPy Library~\cite{Hauschild2018}.
We acknowledge support from the Deutsche Forschungsgemeinschaft (DFG, German Research Foundation) under Germany’s Excellence Strategy--EXC--2111--390814868, TRR80 and DFG grants No. KN1254/1-2 and No. KN1254/2-1, the European Research Council (ERC) under the European Union’s Horizon 2020 research and innovation programme (grant agreement No. 851161), as well as the Munich Quantum Valley, which is supported by the Bavarian state government with funds from the Hightech Agenda Bayern Plus.

\appendix

\section{Luttinger parameter in the symmetric sector} \label{app:K_Jperp}

\begin{figure}[t!]
\centering
\includegraphics[width=0.47\textwidth]{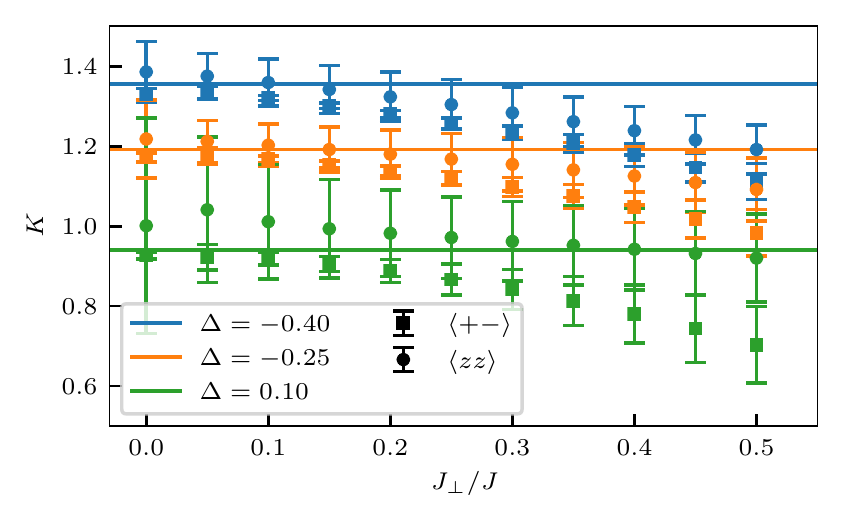}
\caption{Luttinger parameter estimated from a fit to the static correlations in the ground state for a system of size $L=200$ and for different values of the inter chain coupling $J_\perp/J$. The errorbars incorporate the uncertainty of the fit as well as the average over different spatial fitting ranges. The solid lines show the exact result for the single chain as given by Eq.~\eqref{eq:xxz_K}.}   \label{fig:K_correlations}
\end{figure}

As discussed in Section \ref{sec_model}, the bosonization of the coupled chains predicts the emergence of the sine-Gordon field theory in the odd degrees of freedom, while the even sector remains a gapless Luttinger liquid. In this section, we provide a sanity check of this claim by numerically probing the static correlation functions of the ground state and comparing them with their analytic forms given in Refs.~\cite{Hikihara1998,Hikihara2001}.
Indeed, since the contribution of the gapless sector dominates the exponentially-decaying contribution of the massive antisymmetric part $\phi_-$, the critical behavior is already clearly seen in single-chain correlation functions, furthermore giving access to $K$.
The results are shown in Fig.~\ref{fig:K_correlations}. These fits are however hard to perform, particularly in the regime where $\Delta > 0$ due to the large oscillations in the correlation functions, which results  in the large errorbars. We therefore show the results for both
\begin{equation}
\ev{+-}_r= \ev{S^{+}_{\up j}S^{-}_{\up k}}{0} -  \ev{S^{+}_{\up j}S^{-}_{\dw k}}{0}
\end{equation}
and
\begin{equation}
\ev{zz}_r= \ev{S^{z}_{\up j}S^{z}_{\up k}}{0} +  \ev{S^{z}_{\up j}S^{z}_{\dw k}}{0}
\end{equation}
with $r=|j-k|$. Notice that we include the interchain correlations as well, although we are considering the weak coupling limit. We have found that by including them the fits were improved, particularly for $\ev{zz}_r$.  The errorbars shown in the figure are a combination of the uncertainty on the fit, and of the averaging over some different fitting domains $[r_{\mathrm{min}},r_{\mathrm{max}}]$. We incorporate the latter because the fit of $\ev{zz}_r$ is quite sensitive to the choice of $r_{\mathrm{min}}$, due to non-universal short ranged corrections. 

These estimates confirm that the Luttinger parameter of our model agrees with the single chain exact result~\eqref{eq:xxz_K} in the weak tunnelling regime. In particular, deviations are only found around $J_{\perp}/J \gtrsim 0.3$. For our analysis in the main text, we have considered $J_{\perp}/J=0.1,0.2$.

\section{The sine-Gordon scattering matrix and wave-packet displacement}
\label{app_scatmat}
We report the sine-Gordon scattering matrix and its connection with the spatial displacement after a scattering event has occurred. The exact expression of the scattering matrix can be found in Ref.~\cite{Zamolodchikov1979}. From that the breather-breather scattering matrix can be obtained by the form factor bootstrap.
Since we are only interested in breather-breather scattering events, we focus on the latter for simplicity. Let $S^{(n,m)}(\theta)$ be the scattering matrix of a scattering between a breather of species $n$ and $m$, with relative rapidity $\theta$, which for $n\ge m$ reads \cite{mussardo2010statistical}:
\begin{multline}
S^{(n,m)}(\theta)=\frac{\sinh\theta-i\sin\left(\frac{n+m}{2\xi^{-1}}\right)}{\sinh\theta+i\sin\left(\frac{n+m}{2\xi^{-1}}\right)}\frac{\sinh\theta-i\sin\left(\frac{n-m}{2\xi^{-1}}\right)}{\sinh\theta+i\sin\left(\frac{n-m}{2\xi^{-1}}\right)}\times \\
\prod_{k=1}^{m-1}\frac{\sin^2\left(\frac{m-n-2k}{4\xi^{-1}}-i\theta/2\right)}{\sin^2\left(\frac{m-n-2k}{4\xi^{-1}}+i\theta/2\right)}\frac{\cos^2\left(\frac{m+n-2k}{4\xi^{-1}}-i\theta/2\right)}{\cos^2\left(\frac{m+n-2k}{4\xi^{-1}}+i\theta/2\right)}\, ,
\end{multline}
\begin{figure}[t!]
\centering
\includegraphics[width=0.46\textwidth]{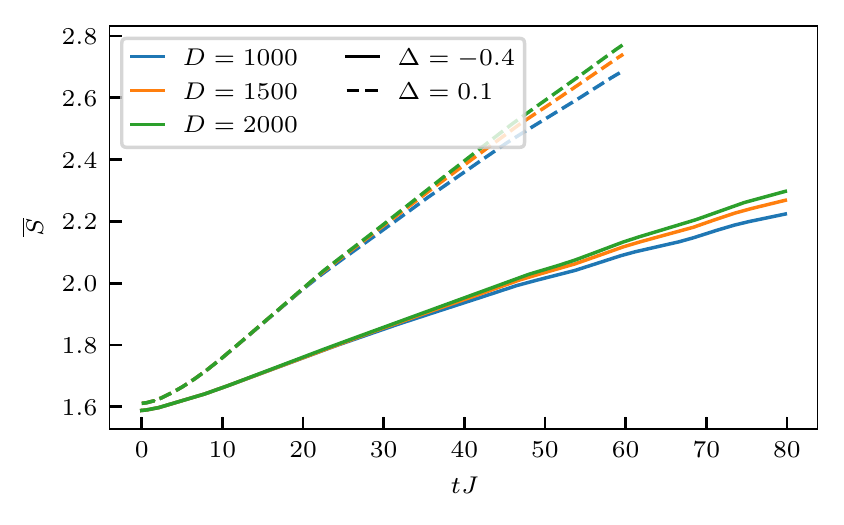}
\caption{The growth of the average entanglement entropy per bond in the time-evolved state $e^{-iHt} S^{+}_{\up j}S^{-}_{\dw j}\ket{0}$ for different bond dimensions $D=1000,1500,2000$ for a ladder system of size $L=151$ and two different $\Delta$. The tunnel coupling is $J_{\perp}/J=0.1$}\label{fig:Saverage_t}
\end{figure}
where the renormalized interaction $\xi$ has been defined below Eq. \eqref{eq:soliton_mass}. The case $n<m$ can be recovered by using the unitary relation $S^{(n,m)}(\theta)S^{(m,n)}(-\theta)=1$.
The scattering matrix is a complex number with modulus $1$, therefore one can define the scattering phase $\Theta^{(n,m)}(\theta)=-i\log S^{(n,m)}(\theta)$.
The derivative of the scattering phase is connected with the spatial displacement after scattering, see e.g. Ref.
\cite{Doyon2018}.
In particular, let $\delta x_{B_n}^{(n,m)}(\theta_n,\theta_m)$ be the spatial displacement experienced by a breather of species $n$ with rapidity $\theta_n$ colliding with a breather of species $m$ and rapidity $\theta_m$, then
\be
\delta x_{B_n}^{(n,m)}(\theta_n,\theta_m)=\frac{1}{cM_{B_n} \cosh\theta_n}\partial_\theta \Theta^{(n,m)}(\theta)\Big|_{\theta=\theta_n-\theta_m}\, .
\ee

This is the analytical expression of the spatial displacement used in Section \ref{sec_scattering}.

\begin{figure*}[t]
\includegraphics[width=0.98\textwidth]{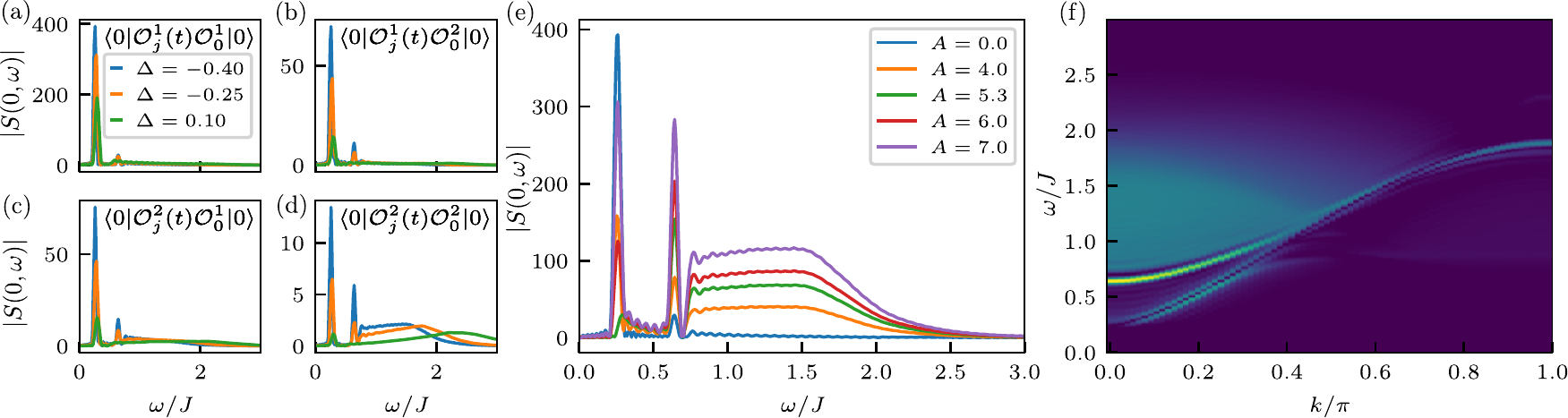}
\caption{(a-d) Spectral functions at $k=0$ evaluated from the correlation functions given in the panels for two different values of $\Delta$. The interchain coupling is $J_{\perp}/J=0.1$, and the system size $L=151$. The peaks correspond to $B_1$ and $B_3$ and their relative strengths can be compared. For the two rung operators there is a continuum present, see panel (d). (e) Spectral functions from the combination of all contributions shown in (a-d) for $\Delta=-0.4$ according to Eq.~\eqref{eq:corr_contr} at $k=0$ for some different values of $A$. The $B_1$ peak is reduced the most for $A=5.3$. (f) Spectral function for $A=5.3$. Here $\Delta=-0.4$, the interchain coupling is $J_{\perp}/J=0.1$ and the system size is $L=151$.}\label{fig:reduce_B3}
\end{figure*}

\section{Technical details on the calculation of spectral functions}\label{app:det_spectral}
We provide some technical aspects on our matrix-product state simulations for the spectral functions shown in Sec.~\ref{sec_spectralF}. Our evaluation of the spectral function~\eqref{eq:spectral_f} relies on the computation of the unequal-time correlation function $C_j(t) = \ev{O_j^{\dag}(t)O_0}{0}$.  
We explicitly perform the real-space time evolution and then the Fourier transformations to momentum and frequency space.
To obtain the time evolution, we apply the TEBD algorithm ~\cite{Vidal2004,White2004,Daley2004}  to the state $O_0\ket{0}$, were $\ket{0}$ is the ground state of the ladder~\eqref{eq:H} obtained from DMRG~\cite{White192,Schollwock2011}. We group the sites connecting each rung of the ladder (labelled by $\up,\dw$) to a single site, and map the ladder system to a chain with an increased local Hilbert space of dimension $d=4$. Within this chain there are only two-site interactions. Hence, we can apply a fourth order TEBD scheme efficiently. As a next step, it suffices to compute the overlaps with $O_j\ket{0}$ at different time steps as
\begin{equation}
C_j(t) = e^{iE_0 t}  \ev{O_j^{\dag} e^{-iHt} O_0}{0}
\end{equation}
with $E_0$ the ground-state energy. The operator $O_0$ acts in the middle of the system and ideally we want to exclude boundary reflections. Therefore, the maximal time of the simulation is chosen as $t_{\mathrm{max}} \lesssim L/(2v_s)$ with $v_s$ the sound velocity. As the resolution in frequency space is directly related to $t_{\textrm{max}}$, we simulate the largest system sizes possible. In practice, we simulated ladders up to $L=151$ rungs for moderate $\Delta \in \left[-0.4,0.1\right]$. The sound velocity is set by $\Delta$ (see Eq.~\eqref{eq:xxz_K} in the main text), so also $t_{\mathrm{max}}$ and the frequency resolution will depend on $\Delta$. As the sound velocity becomes larger with increasing $\Delta$, the entanglement growth will be also more pronounced. In Fig.~\ref{fig:Saverage_t} we show the entanglement growth in a chain of $L=151$ grouped sites, with $J_{\perp}/J=0.1$, for two different values of $\Delta$ each at three different bond dimensions, where
\begin{equation}
\overline{S} = \frac{1}{L-1} \sum_b S_{\textrm{vN}}(b) \; \textrm{with} \quad S_{\textrm{vN}}(b) = -\Tr{\rho_b \log \rho_b}.
\end{equation}
with $b$ running over all bonds, and $\rho_b$ the bipartite reduced density matrix with respect to that bond. In the simulations shown in the main text, we typically use a bond dimensions of $D = 1500$.
Before taking the Fourier transform to frequency space, we multiply the data with a Gaussian window with standard deviation $\sigma$~\cite{White2004}, which we tune as function of the maximal time of the simulation $t_{\mathrm{max}}$. This sets the resolution in frequency space, and induces a broadening of the peaks.

\section{Form factor suppression of $B_1$} \label{app:suppr_B1}

The relative spectral weight of the breathers in the spectral function can be tuned by considering a linear combination of a single and a double rung operator, instead of just the single rung operator. This could be useful because the spectral weight of the breathers $B_n$ with breather index $n>2$ is strongly suppressed compared to the lower lying ones $B_1,B_2$. Here we discuss a sine-Gordon form factor argument to reduce the weight of $B_1$ compared to $B_3$ (we will work in the odd breather sector). Introducing the single and two rung operators 
\begin{align}
\begin{split}
O^{1}_j &= S^{+}_{j\up }S^{-}_{j\dw } \\
O^{2}_j &= S^{+}_{j\up }S^{+}_{j+1\up}S^{-}_{j\dw }S^{-}_{j+1 \dw},
\end{split}
\end{align}
which according to bosonization are ($\alpha_{1,2} \in \mathbb{R}$)
\begin{equation}
O^{1} = \alpha_1 e^{i\phi_-}, \quad O^{2} = \alpha_2 e^{2i\phi_-}.
\end{equation}
Our goal is to find a linear combination $\tilde{O}=O^{1} + AO^{2}$ of these operators such that
\begin{equation}\label{eq:FF_eliminateB1}
\mel{0}{\tilde{O}}{B_1} \simeq  0.
\end{equation}
Within sine-Gordon, we have the following form factor for $B_1 - 0$ 
\begin{equation}
\mel{0}{e^{im\phi_-}}{B_1} = 2i\ev{e^{im\phi_-}}{0}\frac{\sin(m\pi\xi)}{\sin(\pi\xi)}\sqrt{\frac{\sin(\pi\xi)}{F_{\mathrm{min}}}}
\end{equation}
with 
\begin{equation}
F_{\mathrm{min}} = \frac{1}{\cos(\pi\xi/2)}\exp\left( \frac{1}{\pi} \int_0^{\pi\xi} \dd t\frac{t}{\sin t} \right). 
\end{equation}
We have that 
\begin{align}
\begin{split}
\ev{O^{1}}{0} &= \alpha_1 \ev{e^{i\phi_-}}{0}, \\
\ev{O^{2}}{0} &= \alpha_2 \ev{e^{2i\phi_-}}{0},
\end{split}
\end{align}
and therefore to satisfy Eq.~\eqref{eq:FF_eliminateB1} 
\begin{equation}\label{eq:no_B3}
A \simeq -\frac{1}{2\cos(\pi\xi)} \frac{\ev{O^{1}}{0}}{\ev{O^{2}}{0}}.
\end{equation}
Due to the interchain symmetry $\ev{O^{1,2}}{0} = \ev{O^{1,2}}{0}^{*}$, we can simply translate the argument to the anti-symmetric combination of the operators 
\begin{align}
\begin{split}
\mathcal{O}^{1,2}_j &= iO^{1,2}_j + \mathrm{h.c.} \\
\tilde{\mathcal{O}}_j &=\mathcal{O}^{1}_j + A\mathcal{O}^{2}_j.
\end{split}
\end{align}
Such that when we compute the spectral function Eq.~\eqref{eq:spectral_f} with $O_j(t) = \tilde{\mathcal{O}}_j(t)$, $B_2$ vanishes because of parity, and $B_1$ is significantly reduced by the above construction. We have numerically computed the four contributions to the correlation function separately
\begin{multline}\label{eq:corr_contr}
\ev{ \tilde{\mathcal{O}}_j(t) \tilde{\mathcal{O}}_0 }{0} =  \ev{ \mathcal{O}^1_j(t) \mathcal{O}^1_0}{0} + A \ev{ \mathcal{O}^1_j(t) \mathcal{O}^2_0}{0} \\
+ A \ev{ \mathcal{O}^2_j(t) \mathcal{O}^1_0}{0} 
+ A^2 \ev{ \mathcal{O}^2_j(t) \mathcal{O}^2_0}{0}.
\end{multline}
\begin{figure}
\centering
\includegraphics[width=0.49\textwidth]{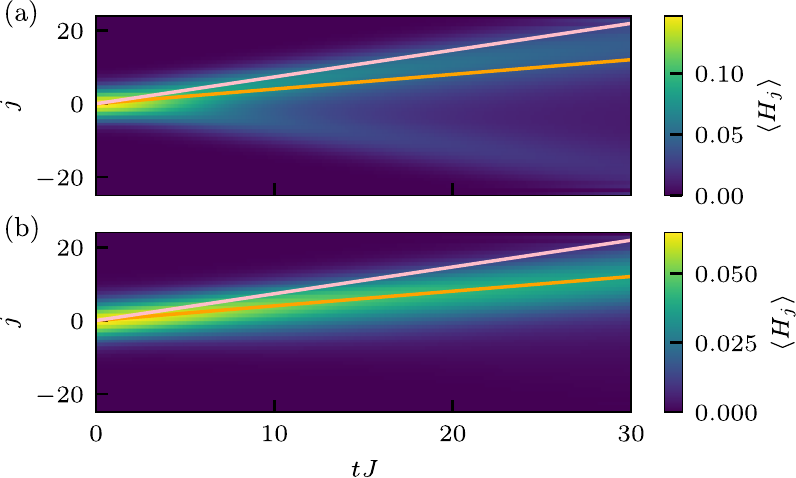}
\caption{The real-time evolution of the state $e^{-\tau H} O^{(2)}_{0,k_2}\ket{0}$. (a) Without imaginary time evolution, $\tau=0$. (b) With imaginary time evolution, $\tau=4/J$. The wave packet is centred initially around momentum $k_2=0.15\pi$ and its width is $\bar \sigma=4$. The system size is $L=51$ and the model parameters are $\Delta=-0.4$ and $J_{\perp}=0.1J$. In the case without imaginary time evolution, there is a two-particle contribution beyond sine-Gordon, that moves with a velocity close to the sound velocity $v_s$ (pink line). A short imaginary time evolution eliminates this high-energy contribution, and we recover a single mode which moves with a velocity compatible with the group velocity as read off from the $B_2$ dispersion (orange line). }\label{fig:L51_e_wp}
\end{figure}
Their respective spectral functions at $k=0$ are shown in Fig.~\ref{fig:reduce_B3} for different values of $\Delta$. Each of the spectral functions detects $B_1$ and $B_3$, however, with different relative weights. In the spectral function obtained from the two-rung operators, Fig.~\ref{fig:reduce_B3}(d), there is also a  continuum contribution present. When summing up the correlation functions according to Eq.~\eqref{eq:corr_contr} we can indeed reduce the contribution of the first peak significantly. The resulting spectral function at $k=0$ are shown in Fig.~\ref{fig:reduce_B3} for different values of $A$, with $\Delta = -0.4$ and $J_{\perp}/J=0.1$. Making $A$ larger will always increase the spectral weight of the continuum, however the relative weight between $B_1$ and $B_3$ reaches a minimum at some intermediate $A$, that we have found numerically to be around $A=5.3$. This can be compared to the value obtained from Eq.~\eqref{eq:no_B3} by filling in the ground state expectation values (in the bulk) and the single-chain $K$, $A_{\text{Eq.~\eqref{eq:no_B3}}}\approx 6.3$. The two values are in decent agreement.  

\begin{figure}[t]
\centering
\includegraphics[width=0.45\textwidth]{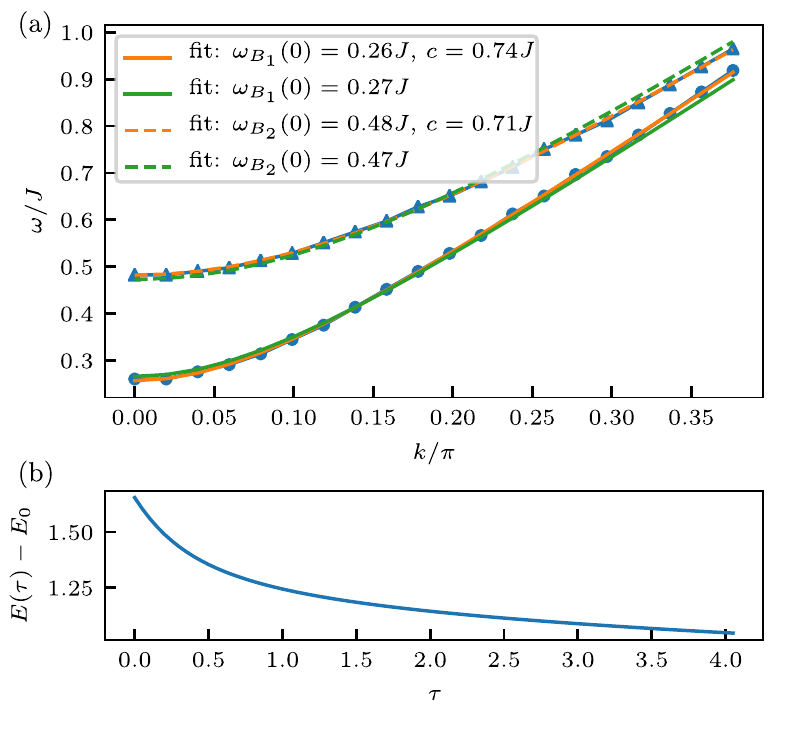}
\caption{(a) Low momenta fits of the breather branches $B_1$ and $B_2$. The data extracted from the spectral function is shown by the circles ($B_1$) and triangles ($B_2$). The lines are fits according to the massive relativistic dispersion $\omega_{B_n}(k)=\sqrt{k^2c^2+M_{B_n}^2c^4}$ with either $c$ and $\omega_{B_n}(0)=M_{B_n} c^2$ as a fitting parameters (orange lines) or with only the latter as free parameter with $c=v_s$ given by Eq.~\eqref{eq:xxz_K} (green lines). 
(b) The energy with respect to the ground state energy of the state $e^{-\tau H} O^{(2)}_{-d,k_2}O^{(1)}_{d,k_1}\ket{0}$ as a function of imaginary time. We stop the imaginary time evolution when $E(\tau)-E_0 \lesssim \omega_{B_1}(k_1)+\omega_{B_2}(k_2)$.  
The system size is $L=101$ and the model parameters are $\Delta=-0.4$ and $J_{\perp}=0.1J$.
}\label{fig:L01_spectum_imagt}
\end{figure}

\section{Technical details on the creation of wave packets}\label{app:det_wp}

The action of the operators given in Eqs.~\eqref{eq:wp_odd} and \eqref{eq:wp_even} on the ground state does not only excite the low-energy sine-Gordon part of the spectrum, but also higher-energy contributions. Indeed, in Fig.~\ref{fig:L51_e_wp}(a), we show the time evolution of the state $O^{(2)}_{0,k_2}\ket{0}$ with $k_2=0.15\pi$ and $\ket{0}$ the ground state. There is a highly-energetic two-particle contribution, as there is also an outgoing signal with negative momentum, that blurs the signal of the sine-Gordon breather. To eliminate this high-energy continuum contribution, we `cool down' the system by performing a short imaginary-time evolution $e^{-\tau H}$, maximally up to imaginary times of around $\tau \approx 4/J$. After this, we indeed obtain a clear single mode that is moving with a speed lower than the sound velocity and compatible with the group velocity $v(k)=\partial_k E_{B_2}(k)$ according to the massive-relativistic dispersion of the second breather, as illustrated in Fig.~\ref{fig:L51_e_wp}(b). 

We estimate the group velocities of our wave packets from fitting the massive-relativistic dispersion relation to the individual breather dispersions as obtained from the spectral functions. We restrict those fits to small momenta. We find that the fitted $c$ of $B_1$ and $B_2$ slightly differs, as shown in Fig.~\ref{fig:L01_spectum_imagt}(a). To enforce consistency, we fixed $c$ to the single chain result of the sound velocity $v_s$, given by Eq.~\eqref{eq:xxz_K}. For $\Delta=-0.4$, this corresponds to $v_s=0.726J$. This value is also approximately the average of the fitted $c$'s shown in the legend of Fig.~\ref{fig:L01_spectum_imagt}(a). We obtain the breather mass from the fitted value of the gap via $M_{B_n}=\omega_{B_n}(0)/c^2$, as indicated by the green lines.

There are no conserved quantities during the imaginary time evolution to cool down the state (this is always the case with TEBD-based algorithms). In addition, as imaginary time evolution is non-unitary the energy will not be conserved. In the long-time limit, imaginary time evolution acts as a projector on the ground state. However, we are not interested in the long time limit, and start from a state that is orthogonal to the ground state. By choosing the imaginary time-step small enough, we used $d\tau=0.0025J$ in our simulations, the state stays orthogonal to the ground state (at least up to the relevant time scales). After the short imaginary time evolution, we checked that (i) the obtained state is still orthogonal to the ground state, and (ii) that the energy $E(\tau)$ of the obtained state is roughly compatible with $E(\tau)-E_0 \approx \omega_{B_1}(k_1)+\omega_{B_2}(k_2)$ with $E_0$ the ground state energy. The total energy difference as a function of imaginary time is shown in Fig.~\ref{fig:L01_spectum_imagt}(b). The imaginary time evolution can in principle also lower the energy of the excitation along the single-particle breather dispersion, such that the group velocity of the wave packet could be overestimated when obtained from the initial momenta that are put in Eqs.~\eqref{eq:wp_odd},\eqref{eq:wp_even}. We have checked this for our data, and this effect seems to be present but not very strong. When a higher precision is required, we directly fitted the velocity of our wave packets instead of determining it via the input momentum.

\bibliography{biblio}

\end{document}